\documentclass[letterpaper]{JHEP3}
\usepackage{amsmath, amsfonts}
\usepackage{epsfig}
\usepackage{cite}

\def\be{\begin{equation}}
\def\ee{\end{equation}}
\def\Mrest{M_{\rm rest}}
\def\Mkin{M_{\rm kin}}

\preprint{{\tt hep-th/0605191}}

\title{Energy Loss of a Heavy Quark from Asymptotically $AdS$ Geometries}

\author{Christopher~P.~Herzog
\\
Department of Physics, University of Washington, 
Seattle, WA 98195-1560 \\
\email{herzog@phys.washington.edu}
}

\abstract{
We investigate some universal features of 
AdS/CFT models of heavy quark energy loss.
 In addition, as a specific example, we examine quark damping in 
 the spinning D3-brane
solution dual to ${\mathcal N}=4$ $SU(N)$ 
super Yang-Mills at finite temperature and
R-charge chemical potential.
}


\begin{document}

\section{Introduction}

We continue the program begun in \cite{HKKKY} of using the AdS/CFT correspondence \cite{jthroat, GKP, EW}
to model the energy dissipation of a heavy quark moving through a plasma.\footnote{
  Recently, several other papers discussing quark damping from other AdS/CFT perspectives have    
  appeared \cite{Rajagopal, SolanaTeaney, Gubser, Buchelnew, Caceres}.}
Damping of heavy quarks is interesting experimentally for understanding charm and bottom 
physics at RHIC \cite{Adler:2005xv,
CalderondelaBarcaSanchez:2005yp,
Suaide:2005yq}.  
The traditional field theoretic approach to the problem is perturbative,
assuming that the quark interacts weakly with the surrounding plasma via two-body collisions with thermal quarks and gluons and via gluon bremsstrahlung  
(see \cite{HKKKY} for a list of relevant papers).  
AdS/CFT provides a dual model where one can calculate the energy dissipation at strong coupling, a regime potentially more interesting for RHIC physics where the effective $\alpha_s$ is believed 
to be of order one.

A direct comparison of AdS/CFT and RHIC data is fraught with difficulty.  The essential problem
is that AdS/CFT does not provide a dual model of QCD with three flavors.
Instead, the original correspondence provides a duality between the maximally 
supersymmetric ${\mathcal N}=4$ $SU(N)$ Yang-Mills theory and type IIB string theory in a $AdS_5 \times S^5$ background.  String theory in curved space 
is not under good theoretical control, but in the limit 
$N \to \infty$ and 
of large 't Hooft coupling  $\lambda = g_{YM}^2 N$, the string theory is well approximated 
by classical supergravity, and strong coupling calculations on the Yang-Mills side reduce to classical calculations in general relativity.

Since the introduction of the original AdS/CFT correspondence, 
a number of generalizations have been made, some of which are 
relevant for our discussion.  By adding a black hole to the supergravity description, the field theory is raised to a finite temperature \cite{Witten:1998zw} dual to the Hawking temperature of the black hole.  
By introducing black holes spinning in the $S^5$ direction, the field theory is raised to a finite chemical potential associated to the R-charges of the supersymmetry algebra \cite{Gubser:1998jb, Behrndt, Cai:1998ji, Chamblin:1999tk,  CveticGubser}.
More recently, \cite{KarchKatz} argued that adding D7-branes to the geometry was dual
to adding flavor hypermultiplets to the gauge theory.

The result of these refinements is a strongly coupled field theory in the large $N$ and $\lambda$ limit with ${\mathcal N}=2$ supersymmetry, a field theory with markedly different field content and interactions than three flavor $N=3$ QCD.  Nevertheless, at finite temperature, the theories may not be so different.  For example, the pressure divided by the free Stefan-Boltzmann limit (which effectively just counts the number of degrees of freedom) in ${\mathcal N}=4$ SYM is remarkably close to the corresponding ratio in QCD at temperatures of a few times $T_c$ where it is strongly coupled \cite{GubserF}.  The dimensionless ratio of viscosity divided by entropy density
equals $1/4\pi$ in ${\mathcal N}=4$ SYM, as well as in all other theories
with gravity duals \cite{kovtun, Buchel} in the strong 't Hooft coupling
limit.
And this value, which is lower than any weakly coupled theory
or known material substance \cite{Kovtun:2004de}, is in good agreement with
hydrodynamic modeling of RHIC collisions \cite {Shuryak,Shuryak:2004cy}.

Although one could introduce a chemical potential for heavy quarks, a $U(1)_R$ potential
is a better way of capturing effects analogous to a light quark chemical potential.
For heavy quarks, introducing a baryon number chemical potential in the AdS/CFT context would mean introducing a macroscopic density of heavy quark baryons.  More physically relevant is a situation with a few heavy quark probes moving through a soup with a density of ordinary baryons made up of lighter quarks.

The lesson we draw from the successful comparisons of pressure and viscosity is that 
we should try to search out dimensionless ratios and also 
universal features of dual AdS/CFT models.  To this end, we take a more general perspective
than in \cite{HKKKY} where attention was restricted to the dual
model of finite temperature ${\mathcal N}=4$ SYM with one ${\mathcal N}=2$ hypermultiplet. 
We consider a general form for the metric of the gravity dual which has a horizon, is asymptotically
$AdS_{d+1}$ and preserves Poincare invariance on the boundary.  Such a metric includes
the finite temperature, zero chemical potential case 
studied in \cite{HKKKY} as well as the finite $U(1)_R$ chemical potential case mentioned above.  
Such a metric ought also to include a number of relevant deformations of SYM,
and other more speculative AdS/CFT correspondences in dimensions $d \neq 4$.

To this $d+1$ dimensional metric, we assume a flavor brane can be added
as in \cite{KarchKatz}.  This flavor brane should fill all of the asymptotically $AdS_{d+1}$ 
space down to some minimal radius $u=u_0$.  There will in general be a nontrivial relation
between the radius $u_0$ and the Lagrangian mass $m$ of the quark.  We model our heavy quark as
 a classical string that stretches from the flavor brane to the horizon of the black hole.  The rest mass
 $\Mrest$
 of the quark is the energy of a straight, motionless string stretching from 
 $u=u_0$ to the horizon $u=u_h$ and will in general be related in a nontrivial way to both $m$ and $u_h$.  In the limit $u_0$ becomes large compared to the other scales in the problem,\footnote{
 At the very least, $u_0 \gg u_h$, but in general there may be other scales.} 
 $\Mrest$ and $m$ scale linearly with $u_0$.
 
 There is a substantial amount of interesting physics in the relation between $u_0$, $m$, and $\Mrest$
 which we will for the most part ignore in this paper.  It is known that 
 in the case dual to ${\mathcal N}=4$ SYM at finite temperature and R-charge chemical potential, 
 there is a first order phase transition 
 as $u_0$ is lowered toward the horizon.  For example, in the case of finite temperature and zero potential, 
 at a value $u_0 / u_h \approx 1.02$, the flavor
 brane jumps to a configuration where it intersects the horizon \cite{Babington:2003vm,andy,Mateos:2006nu,Albash:2006ew}.  A similar jump occurs at finite potential \cite{AlbashRcharge} and perhaps  
 may happen more generally.  There are two lessons to keep in mind.  
 One is that in the limit where the radius
 of curvature $L$ of $AdS$ is kept large, the string will remain classical down to the phase transition
 point.  The second is that the configuration $u_0=u_h$ is likely unstable.
 
 Although our approach is more general, the tools we use to measure energy dissipation of a heavy 
 quark are similar to \cite{HKKKY}.  The goal is to calculate the friction coefficient $\mu$ in the equation
 \be
 \frac{dP}{dt} = -\mu P \ .
 \ee
 In Section \ref{sec:analytic}, we revisit the analytic solution of a string moving at constant velocity $v$.  Such a string is dual to a quark in an external electric field, and we are  
 able to extract the amount of 
 momentum and energy the field must supply to keep the quark in motion.
 In Section \ref{sec:linear}, we revisit the linearized, quasinormal mode analysis of the string equation of motion. 
 The quasinormal modes give information about the return to equilibrium of the string after small perturbations, and thus tell us about $\mu$ in the small $v$ limit.  In Section \ref{sec:Rcharge},
 we consider a specific example, the R-charge black hole dual to the SYM field theory at finite
 temperature and $U(1)_R$ chemical potential.
 
Saving the details for the body of the paper, we make three interesting observations about
our results.  The first concerns the small $v$ limit of $\mu$.  Both the
analytic, constant velocity solution and the quasinormal mode analysis confirm that for all the cases considered
\be
\mu \Mkin = \frac{ (4 G_N s)^{2/(d-1)} }{2 \pi \alpha'}
\label{largemmu}
\ee
 where $1/(2\pi \alpha')$ is the string tension, 
 $G_N$ is Newton's constant, and $s$ is the entropy density.  
 We have introduced a new mass, $\Mkin$, the kinetic mass which enters into the dispersion relation for the quark.  In the large $u_0$ limit, we expect $\Mkin \sim \Mrest$.
 In the case of asymptotically 
 $AdS_5$ geometries, the dual field theory should be a variation of
 ${\mathcal N}=4$ $SU(N)$ 
 super Yang-Mills.  In this case $G_N = \pi L^3 / 2 N^2$ where $N$ is the number of colors in the 
 associated $SU(N)$ field theory.  Also, $L^2 / \alpha' = \sqrt{\lambda}$ where $\lambda = g_{YM}^2 N$
is the 't Hooft coupling.  We find that in this case, (\ref{largemmu}) becomes
\be
\mu \Mkin = \sqrt{\lambda} \left( \frac{s^2}{2 \pi N^4} \right)^{1/3} \ .
\ee

The second observation concerns the small $u_0$ limit.  Although the D7-brane with 
$u_0 \approx u_h$
is probably not stable, and a string stretching from such a D7-brane to the horizon is
more quantum than classical, it is nevertheless possible to analyze the quasinormal mode problem
formally in this limit:
\be
\mu = 2 \pi T\ .
\ee
Moreover, in the R-charge black hole case analyzed in Section \ref{sec:Rcharge}, $\mu$ is a monotone
decreasing function of $u_0$, leading us to speculate that $\mu$ is always bounded above by 
$2 \pi T$ in the non-relativistic regime.

The last observation concerns the velocity dependence of $\mu$.  In \cite{HKKKY}, limited
evidence supported the claim that the heavy quark obeys a relativistic dispersion relation
\be
E(P) = \Mrest - \Mkin + \sqrt{P^2 + \Mkin^2} \ .
\ee
Moreover, in \cite{HKKKY}, the friction coefficient $\mu$ was velocity independent.
Assuming that the same dispersion relation holds in the case of finite chemical potential studied in Section \ref{sec:Rcharge}, we are able to extract $\mu \Mkin$ from the constant velocity solution of Section \ref{sec:analytic}.  Our results indicate that $\mu$ 
has a strong velocity dependence, increasing as $v$ increases.  
Notably, the perturbative calculations of quark damping also have a nontrivial velocity dependence (see for example \cite{MooreTeaney}).

\section{The equations of motion}
\label{sec:setup}

We assume a metric of the form
\be
ds^2 = g_{tt} dt^2 + g_{uu} du^2 + g_{xx} \delta_{ij} dx^i dx^j \ ,
\ee
where $i = 1, 2, \ldots, d-1$.
As $u \to \infty$, the metric should approach that of $AdS_{d+1}$ with a radius
of curvature $L$:
\be
g_{tt} \to -L^2 u^2  \; ; \; \; \;
g_{uu} \to  \frac{L^2}{u^2} \; ; \; \; \;
g_{xx} \to L^2 u^2 \ .
\label{asympt}
\ee
The space is also assumed to have a horizon at $u = u_h$:
\be
g_{tt} = (u-u_h) \partial_u g_{tt}(u_h) + \ldots \; ; \; \; \;
g^{uu} = (u-u_h) \partial_u g^{uu}(u_h)+ \ldots \ .
\ee
The metric component $g_{xx}$ is assumed to be finite at $u=u_h$.
Finally, we assume that the metric components $g_{tt}$, $g_{uu}$, and $g_{xx}$ depend only
on the radial coordinate $u$.  As a shorthand, we will take $\partial_u f = f'$ and $\partial_t f = \dot f$.

As discussed in the Introduction, this metric includes as special cases a wide variety of space-times dual, via the AdS/CFT correspondence, to strongly coupled field theories.  Some examples are
finite temperature ${\mathcal N}=4$ $SU(N)$ SYM in $d=4$ dimensions
discussed in \cite{HKKKY}, the
same finite temperature SYM at finite R-charge chemical potential to be discussed in 
Section \ref{sec:Rcharge}, various relevant deformations of SYM, and other more speculative AdS/CFT correspondences in $d \neq 4$.

The Hawking temperature of this black hole space-time, dual via the AdS/CFT dictionary to the
temperature of the field theory, 
can be computed by checking that the Euclidean continuation of the metric
is regular at $u=u_h$.
In this case, we find that
\be
T = \left. \frac{\sqrt{-(g_{tt})'(g^{uu})'}}{4 \pi} \right|_{u=u_h}\ .
\ee
The entropy density of the field theory, proportional to the area of the black hole, is
\be
s = \left. \frac{(g_{xx})^{(d-1)/2}}{4 G_N} \right|_{u=u_h} \ .
\ee

We model a quark in the field theory as a classical string in the dual space-time.
We derive the equations of motion for the string from the Nambu-Goto action
\be
S = -\frac{1}{2\pi \alpha'} \int d\sigma \, d \tau \, \sqrt{-G}
\ee
where $G_{ab}$ is the induced metric on the string world-sheet.  
We take a static gauge where
$t = \tau$, $u = \sigma$, and the string only extends in one direction $x(\sigma, \tau)$.
Defining $X = (t, u, x)$ and $U \cdot V = U^\mu V^\nu g_{\mu\nu}$ where $g_{\mu\nu}$
is the space-time metric, we find
\begin{eqnarray}
-G &=& (\dot X \cdot X')^2 - (X')^2 (\dot X)^2 \nonumber \\
&=& -g_{uu} g_{tt} - g_{xx} g_{tt} (x')^2 - g_{uu} g_{xx} \dot x^2 \ .
\end{eqnarray}
The equation of motion is a partial differential equation:
\be
\partial_u \frac{-g_{xx} g_{tt} x'}{\sqrt{-G}} - g_{uu} g_{xx} \partial_t \frac{\dot x}{\sqrt{-G}} = 0 \ .
\label{eom}
\ee

Recall that the canonical momentum densities associated to the string are
\begin{eqnarray}
\pi_{\mu}^0 &= &
-\frac{1}{2\pi \alpha'} g_{\mu\nu} \frac{(\dot X \cdot X') (X^\nu)' - (X')^2 (\dot X^\nu)}{\sqrt{-G}} \ , \\
\pi_{\mu}^1 &=&
 -\frac{1}{2\pi\alpha'} g_{\mu\nu}  \frac{(\dot X \cdot X') (\dot X^\nu)- (\dot X)^2 ( X^\nu)'}{\sqrt{-G}} \ .
\end{eqnarray}
For our string, these expressions reduce to
\be
\label{pdensity}
\left(
\begin{array}{cc}
\pi^0_x & \pi^1_x \\
\pi^0_u & \pi^1_u\\
\pi^0_t  & \pi^1_t
\end{array} 
\right)
=
\frac{1}{2 \pi \alpha'} \frac{1}{\sqrt{-G}}
\left(
\begin{array}{cc}
 g_{xx} g_{uu} \dot x & g_{xx}g_{tt} x' \\
-g_{uu} g_{xx} \dot x x' & g_{uu} (g_{tt} + g_{xx} \dot x^2) \\
g_{tt} (g_{uu} + g_{xx} (x')^2) & -g_{tt} g_{xx} \dot x x'
\end{array}
\right) \ .
\ee

There is a simple time independent solution to (\ref{eom}), namely $x=A$ where $A$ is a constant
and the string stretches from a D7-brane at $u=u_0$ to the horizon at $u=u_h$.
Let's calculate the total energy of such a configuration:
\be
E = -\int_{u_h}^{u_0}  \pi_t^0 \, du= \frac{1}{2\pi\alpha'} \int_{u_h}^{u_0} \sqrt{-g_{tt} g_{uu}} \, du \ .
\ee
This energy is naturally associated with the rest mass of the quark $E \equiv \Mrest$.
If we take $u_0 \to \infty$, there is a divergence associated to this integral.  The limit $u_0\to u_h$, 
on the other hand, is finite.  Assuming $g_{tt}$ and $g_{uu}$ are well behaved in between,
we may conclude that as $u_0\to \infty$, 
\be
\Mrest \to u_0 L^2 / (2 \pi \alpha') \ .
\ee

\section{An analytic, time dependent solution}
\label{sec:analytic}

Assuming $\partial_t x = v$ with $v$ a constant, we will find an analytic solution of (\ref{eom}) dual
to a single quark moving in an electric field ${\mathcal E}$.
The equation of motion reduces in this case to
\be
\partial_u \frac{-g_{xx} g_{tt} x'}{\sqrt{-G}} = 0
\label{reducedeom}
\ee
where
\be
-G = -g_{uu} g_{tt} - g_{xx} g_{tt} (x')^2 - g_{uu} g_{xx} v^2 \ .
\ee
Integrating once with respect to $u$, (\ref{reducedeom}) transforms into
\be
x' = \frac{\sqrt{-G}}{-g_{xx} g_{tt}} C
\ee
where $C$ is the constant of integration.
Solving now for $x'$ yields
\be
(x')^2 = -\frac{g_{uu} C^2 (g_{tt} + g_{xx} v^2)}{g_{xx} g_{tt} (g_{xx} g_{tt} + C^2 ) } \ .
\ee

With these results for $x'$ and $\dot x$ in hand, we return to the canonical momentum densities (\ref{pdensity}), finding
\be
\pi^1_x = -\frac{1}{2\pi\alpha'} C\; \; \; \; ; \; \; \; \; \pi^1_t = \frac{1}{2\pi\alpha'} v C \ .
\ee
If we have an open string, then this string will gain energy and momentum 
through an endpoint $u_0$
at a rate
\begin{eqnarray}
\frac{dE}{dt} &=& \left.  \pi_t^1 \right|_{u=u_0}= \frac{1}{2\pi\alpha'} C v \ , \nonumber \\
\frac{dP}{dt} &=& \left. -\pi_x^1 \right|_{u=u_0} = \frac{1}{2\pi\alpha'} C \ .
\label{dPdt}
\end{eqnarray}
and lose an equivalent amount of $P$ and $E$ at the other endpoint.\footnote{
We would like to thank C.~Kozcaz, who independently obtained \protect{(\ref{dPdt})},
for collaboration in the early stages of this project.}

Let's specialize to the case where we have a string that stretches from the D-brane
at $u=u_0$ to the horizon $u=u_h$.  
Such a string can be thought of as a single quark moving in an electric field with strength
$\pi^1_x = -{\mathcal E}$.  The electric field
comes from the $U(1)$ gauge field living on the D-brane and has nothing to do with the
$SU(N)$ gauge field of the field theory dual.  
This ${\mathcal E}$ feeds energy and momentum into the 
string at a rate given by (\ref{dPdt}) sufficient to keep the string moving at a constant velocity.

In order for the string to stretch from $u_0$ to $u_h$, $C$ has to
satisfy a special condition.
We know generically that $g_{tt}$ has a zero at $u=u_h$
from which we can conclude that for small $v$, 
$g_{tt} + g_{xx} v^2$ has a zero for some $u=u_c>u_h$.  
Thus in order for $(x')^2$ and $\sqrt{-G}$ to be well defined along the length of the string,
the factor in the denominator $g_{xx} g_{tt} + C^2$ must have a zero at the same location $u=u_c$.
In general an explicit expression for $u_c$ and $C$ may be difficult to find.

Let us assume, for simplicity, that $v$ is small.  In that case, we can find an approximate expression
for $u_c$:
\be
u_c = u_h-\left. \frac{v^2 g_{xx} }{(g_{tt})'} \right|_{u=u_h} + {\mathcal O}(v^4) \ .
\ee
From this approximation, we find that
\be
C = v g_{xx} (u_h) \left(1 + {\mathcal O}(v^2) \right)\ .
\label{Capprox}
\ee
Thus, putting (\ref{dPdt}) and (\ref{Capprox}) together,  we find that
\be
\frac{dP}{dt} = -\frac{1}{2\pi\alpha'} v g_{xx}(u_h)\left(1 + {\mathcal O}(v^2) \right) \ .
\label{analyticresult}
\ee

\section{Linear Analysis}
\label{sec:linear}

In this section we analyze small perturbations of a straight string which stretches from $u=u_0$
to $u=u_h$.  This analysis allows us to investigate the friction coefficient $\mu$ in the non-relativistic limit 
for any quark rest mass $\Mrest$.

Let's look for a solution to (\ref{eom}) where $-G \approx -g_{uu} g_{tt}$.  Let us also assume
a time dependence of the solution that exhibits exponential damping: $\dot x = -\mu x$.  With these two assumptions, the equations of motion become
\be
 \frac{\sqrt{-g_{tt} g^{uu}}}{g_{xx}} \partial_u \sqrt{-g_{tt} g^{uu}} g_{xx} x' = \mu^2 x \ .
\label{lineom}
\ee
We are interested in solutions $x$ with standard D-brane boundary conditions, i.e. 
Neumann boundary conditions at a radius $u=u_0$.  Because of the absorptive nature
of the black hole, we take
``out-going'' boundary conditions at the horizon $u=u_h$ \cite{SonStarinets}.  
To explain ``out-going'', consider
the solution to (\ref{lineom}) close to the horizon.  Near $u=u_h$, (\ref{lineom}) takes
the approximate form
\be
 (4 \pi T)^2 (u-u_h) \partial_u (u-u_h) x' = \mu^2 x \ ,
\ee
which has the two solutions
\be
x = c_1 (u-u_h)^\gamma + c_2 (u-u_h)^{-\gamma} \ ,
\ee
where
\be
\gamma = \frac{\mu}{4 \pi T} \ .
\ee
Out-going boundary conditions means we take $c_1=0$; with the $e^{-\mu t}$ time dependence, 
waves travel into but not out of the event horizon.

As a first step in this linear analysis, 
we make an assumption that will turn out to correspond to 
studying heavy quarks.  In the next section we will consider light quarks, 
and
in Section \ref{sec:numerics}, we will study
the general case for a specific space-time numerically.  

We assume that $\mu$ is small and
accordingly  expand our solution as a power series in $\mu$:
\be
x = x_0 + \mu^2 x_1 + \ldots 
\ee
where now our differential equation ${\mathcal L}x=\mu^2 x$ breaks apart into pieces
${\mathcal L}x_0 = 0$ and ${\mathcal L}x_1 = x_0$.  The only solution for the leading term is
to take $x_0 = A$ where $A$ is a constant.  Solving now for $x_1$ yields
\be
x_1' = \frac{A}{g_{xx} \sqrt{-g_{tt} g^{uu}}} \int_{u_0} \frac{g_{xx}}{\sqrt{-g_{tt} g^{uu}}} du
\label{wint}
\ee
where we have taken the lower bound of integration to be $u_0$ to satisfy the Neumann
boundary conditions.

As a final approximation, we will take $u_0$ to be very large so that near $u=u_0$,
the metric components take the asymptotic form (\ref{asympt}).  Having taken this final limit,
we can approximately evaluate $x_1'(u)$ for $u$ close to the horizon.  
The key to the evaluation is the realization
that the integral (\ref{wint}) will be dominated by its limit behavior near $u=u_h$ and
near $u_0$.  For $u \approx u_h$, 
\be
x_1'(u)  \approx \frac{A}{ 4\pi T g_{xx}(u_h)(u-u_h)} 
\left(-L^2 u_0 + \frac{g_{xx}(u_h)}{4\pi T} \ln(u-u_h) \right)  \ .
\ee
Matching this result onto our definition of out-going boundary conditions yields
a quasinormal mode condition on $\mu$:
\be
\mu = \frac{1}{L^2 u_0} g_{xx}(u_h)  \ .
\ee

We now use this result for the quasinormal mode to find an expression for the momentum
loss. Identifying the endpoint of the string at $u=u_0$ as a quark, the velocity of the quark
obeys the differential equation $\dot v = - \mu v$.
In the large $u_0$ limit, we have that 
the mass is approximately $\Mrest \sim u_0 L^2 / (2 \pi \alpha')$.  Putting these two
facts together, we find that
\be
\frac{dP}{dt} \approx 
 -\frac{u_0 L^2}{2\pi\alpha'} \mu v
 = -\frac{1}{2\pi \alpha'} v g_{xx}(u_h)
\ee
in perfect agreement with the result (\ref{analyticresult}) of the previous section for a slowly
moving heavy quark.

\subsection{Light Quark Limit}

Having found an analytic expression for $\mu$ in the limit $u_0 \to \infty$, we now investigate 
the opposite limit $u_0 \to u_h$.  As discussed in the Introduction, this limit corresponds to
relatively light quarks.  
In Section \ref{sec:setup}, we made some assumptions about the near horizon behavior of the metric components.  To make progress here, we need to make a few
additional assumptions:
\begin{eqnarray}
\sqrt{-g_{tt} g^{uu}} &=& 4 \pi T (u-u_h) \left( 1+ c (u-u_h)  + \ldots \right) \\
g_{xx} &=& g_{xx}(u_h) + g_{xx}'(u_h) (u-u_h) + \ldots \ ,
\end{eqnarray}
where $c$ is a $u$ independent expression that depends on the details of the metric.\footnote{Note that
there may well be exotic cases where the metric does not have a regular power series expansion 
near the horizon, $u \approx u_h$.}  

Assuming a power series solution to (\ref{lineom}) of the form
\be
x(u) = (u-u_h)^{-\gamma} (1 + A (u-u_h) + \ldots) \ ,
\label{impasymp}
\ee
we solve for $A$.  The result is that
\be
A = \gamma \left( c + \frac{g_{xx}'(u_h)/g_{xx}(u_h)}{1-2 \gamma} \right) \ .
\ee

Our power series expression for $x(u)$ satisfies the required out-going boundary conditions at 
the horizon.  We also require Neumann boundary conditions at the flavor brane:
 $x'(u_0) = 0$.  Generically in the limit $u_0 \to u_h$, we
expect the first few terms in the power series expansion for $x(u)$ to be dominant.
To satisfy Neumann boundary conditions, a sufficient condition is the requirement that 
$(u_0-u_h)/(1-2\gamma)$ be
held fixed in the limit $u_0 \to u_h$.   In this way, there is a possibility that the
 second term in the power series expansion for 
$x'(u)$ can cancel the first one at $u=u_0$.
But this condition tells us that the friction coefficient in this limit
must be
\be
\mu = 2 \pi T \ .
\ee

Given that in the limit $u_0 \to \infty$, $\mu$ scales as $T/u_0$, it is tempting to speculate that
$\mu$ is a monotonically decreasing function of $u_0$.  In the example we study in Section \ref{sec:Rcharge}, this monotone behavior holds.  
Given such a monotone behavior, it is tempting to go even further and speculate that
$\mu /T$ is bounded above by $2\pi$ for every AdS/CFT model of quark damping.

\subsection{Dispersion Relations}

Continuing our linear analysis, 
we attempt to establish a relationship between the energy $E$ and momentum $P$ of the
string assuming a time dependence of the form $\dot x = -\mu x$ and that $\dot x$ and $x'$
are small.

Using the equation of motion (\ref{lineom}), we can rewrite the momentum density as
\be
\pi_x^0 = -\frac{\mu}{2\pi\alpha'} \frac{g_{xx} g_{uu} x}{\sqrt{-g_{tt} g_{uu}}} = -\frac{1}{2\pi\alpha'}\frac{1}{\mu}
\partial_u \frac{-g_{tt} g_{xx} x'}{\sqrt{-g_{tt} g_{uu}}} \ .
\ee
The total momentum integral can now be evaluated
\be
P = \int \pi_x^0 \,  du = \left. -\frac{1}{2\pi\alpha'}\frac{1}{\mu}
 \sqrt{\frac{-g_{tt}}{g_{uu}}} g_{xx} x' \right|_{u=u_{min}}^{u=u_0} \ .
\ee
Because of Neumann boundary conditions at the flavor brane, we know $x'(u_0)=0$.
Ideally, we would like to take $u_{min}=u_h$, but there will be a divergence which
we regulate by introducing an infrared cutoff $u_{min} > u_h$. 

We evaluate the energy in a similar fashion.  Now we keep quadratic terms in
the expansion of $\sqrt{-G}$, anticipating a non-relativistic dispersion relation.  
The energy density takes the form
\be
\pi_t^0 = -\frac{1}{2\pi\alpha'} \left[
 \sqrt{-g_{tt} g_{uu}} + \frac{1}{2} g_{xx} \sqrt{\frac{-g_{tt}}{g_{uu}}}  (x')^2
+ \frac{1}{2} g_{xx} \sqrt{\frac{g_{uu}}{-g_{tt}}}  (\dot x)^2 \right] \ .
\ee
Integrating by parts and using the linearized equation of motion yields a simple
expression for the energy
\be
E = -\int \pi_t^0 \, du = \frac{1}{2\pi\alpha'} \int_{u_{min}}^{u_0} \sqrt{-g_{tt} g_{uu}} du
- \left. \frac{1}{4 \pi \alpha'} g_{xx} \sqrt{\frac{-g_{tt}}{g_{uu}}}  x x' \right|_{u=u_{min}}
\ee
where we have used the fact that $x'(u_0) =0$. 
 Using the fact that close to the horizon
\[
x \sim (u-u_h)^{-\mu/4 \pi T} \ ,
\]
and recalling the definition of $\Mrest$, we find that
\be
E = \Mrest + \frac{1}{2} \frac{P^2}{\Mkin} \ .
\ee
where we have defined a kinetic mass
\be
\Mkin \equiv \frac{g_{xx}(u_h)}{(2 \pi \alpha') \mu} \ .
\ee 
In other words, we have found that the quark obeys essentially the usual, non-relativistic dispersion relation for a point particle.  The only difference is that the rest mass is different from the kinetic mass.

\section{An Example: The R-charged black D3-brane background}
\label{sec:Rcharge}

Consider the following asymptotically $AdS_5$ metric with horizon \cite{Behrndt}:
\be
ds^2 = -{\mathcal H}^{-2/3} \frac{(\pi T_0 L)^2}{\rho} f dt^2 + {\mathcal H}^{1/3} \frac{(\pi T_0 L)^2}{\rho} (dx^2 + dy^2 + dz^2) + {\mathcal H}^{1/3} \frac{L^2}{4 f \rho^2} d\rho^2 \ ;
\ee
where
\be
{\mathcal H} = \prod_{i=1}^3 H_i \; ; \; \; \; H_i = 1 + \kappa_i \rho \; ; \; \; \; 
f = {\mathcal H} - \rho^2 \prod_{i=1}^3 (1+ \kappa_i) \ ,
\ee
\be
T_0 = \frac{r_+}{\pi L^2}  \; ; \; \; \; \kappa_i = \frac{q_i}{r_+^2} \ ,
\ee
and with respect to the radial coordinate $u$ of previous sections $\rho = r_+^2 / (u^2 L^4)$.
With this change of variables in mind, note that $u_h = r_+/L^2$.
The Hawking temperature of the black hole solution is
\be
T = \frac{2+\kappa_1 + \kappa_2 + \kappa_3 - \kappa_1 \kappa_2 \kappa_3}{2 \sqrt{(1+\kappa_1)(1+\kappa_2)(1+\kappa_3)} }T_0 \ .
\ee 

This gravitational background is dual to ${\mathcal N}=4$ $SU(N)$ Super Yang-Mills theory
with finite chemical potential for the $U(1)$
R-charges.  
The chemical potentials are related to the $\kappa_i$ via
\be
\phi_i = \pi T_0 \frac{\sqrt{2 \kappa_i}} {1+\kappa_i} \prod_{j=1}^3(1+\kappa_j)^{1/2} \ .
\ee
For convenience, we will express the masses and friction coefficient in terms of the $\kappa_i$ rather than the $\phi_i$.  

The black hole provides a model in which to explore the effects of chemical
potential on quark damping.  We will find two interesting effects.  The first is that
the friction coefficient is not a monotonic function of the chemical potential.  The second is, assuming a
relativistic dispersion relation for the quark,  that the friction coefficient has 
nontrivial velocity dependence, unlike the zero chemical potential case studied in 
\cite{HKKKY}.

\subsection{Moving Quark}

We begin with a discussion of the analytic, single quark solution discussed in Section 
\ref{sec:analytic}.
Formally, from Section \ref{sec:analytic}, we know that $dP/dt = -C / 2 \pi \alpha'$.  It is tempting
to reorganize this information assuming a relativistic, single particle dispersion relation for the
quark
\be
P = \frac{\Mkin v}{\sqrt{1-v^2}} \ .
\ee
In this case, the friction coefficient $\mu$ can be expressed in terms of $C$ as
\be
\frac{\mu\Mkin}{ T^2 \sqrt{\lambda}} = \frac{\sqrt{1-v^2}}{v} \frac{C}{2 \pi T^2 L^2} \ .
\ee

To consider the small velocity limit of this analytic solution, we evaluate $g_{xx}$
on the horizon:
\begin{eqnarray*}
g_{xx}(u_h) &=& (\pi T_0 L)^2 \prod_{i=1}^3 (1+ \kappa_i)^{1/3}\\
&=& 
(2 \pi T L)^2 \frac{\prod_{i=1}^3 (1+\kappa_i)^{4/3}}
{(2+\kappa_1 + \kappa_2 + \kappa_3 - \kappa_1 \kappa_2 \kappa_3)^2} \ .
\end{eqnarray*}
Since $C = v g_{xx}(u_h)$, we find that
\be
\frac{\mu \Mkin }{T^2 \sqrt{\lambda}} = 2 \pi  \frac{\prod_{i=1}^3 (1+\kappa_i)^{4/3}}
{(2+\kappa_1 + \kappa_2 + \kappa_3 - \kappa_1 \kappa_2 \kappa_3)^2}
\ee
which is shown plotted in Figure \ref{largemass}.  Note that the plots are not monotone increasing
functions of the chemical potential $\kappa_i$.  In the case where $\kappa_2=\kappa_3=0$ and only
$\kappa_1$ is dialed, $\mu \Mkin$ reaches a maximum at about $\kappa_1=1$.

\FIGURE{
\centerline{\psfig{figure=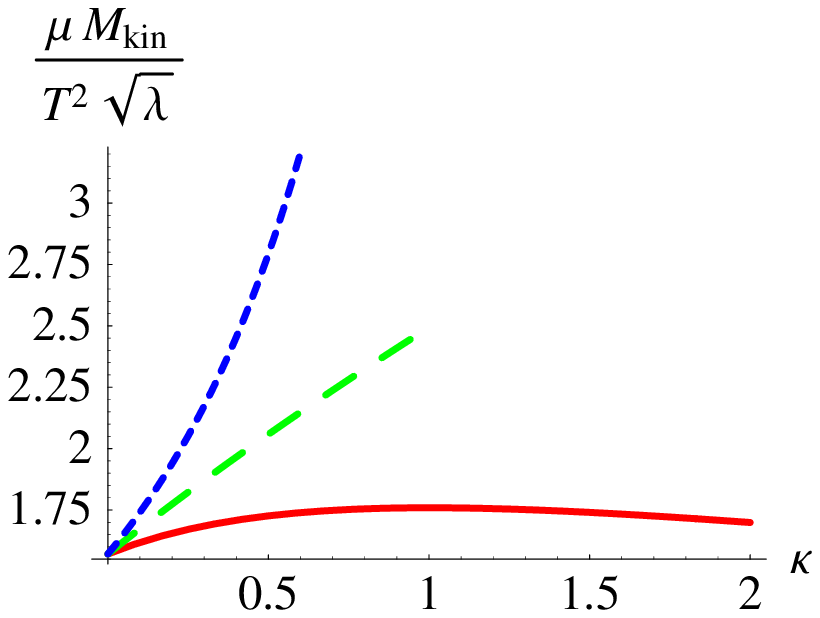, width=3in}}
\caption{The friction coefficient as a function of $\kappa_i$ in the small $v$ limit.  For the solid red line, 
$\kappa_1=\kappa$, $\kappa_2=\kappa_3=0$.  For the dashed green line, 
$\kappa_1=\kappa_2=\kappa$, $\kappa_3=0$.  For the dotted blue line,
$\kappa_1=\kappa_2=\kappa_3=\kappa$.  The domain of the plot is determined
by the region of thermodynamic stability: $2-\kappa_1-\kappa_2-\kappa_3 - \kappa_1 \kappa_2 \kappa_3 > 0$. }
\label{largemass}
}

\FIGURE{
\centerline{\raisebox{3cm}{(a)} \psfig{figure=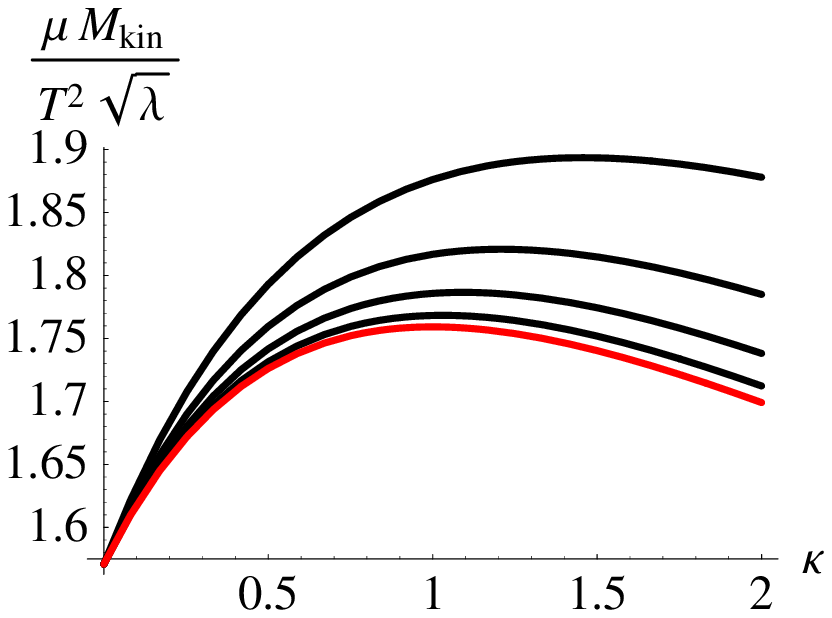, width=3in}
\raisebox{3cm}{(b)} \psfig{figure=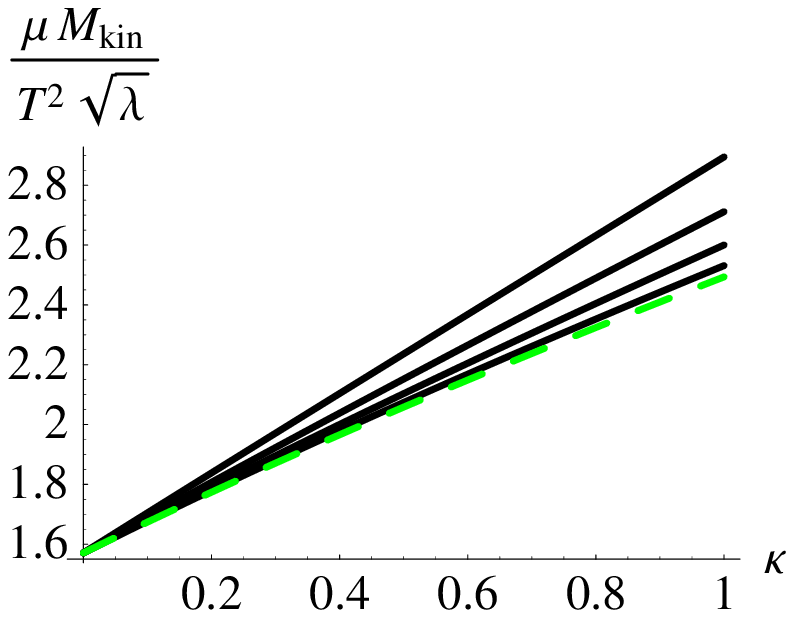, width=3in}}
\centerline{\raisebox{3cm}{(c)} \psfig{figure=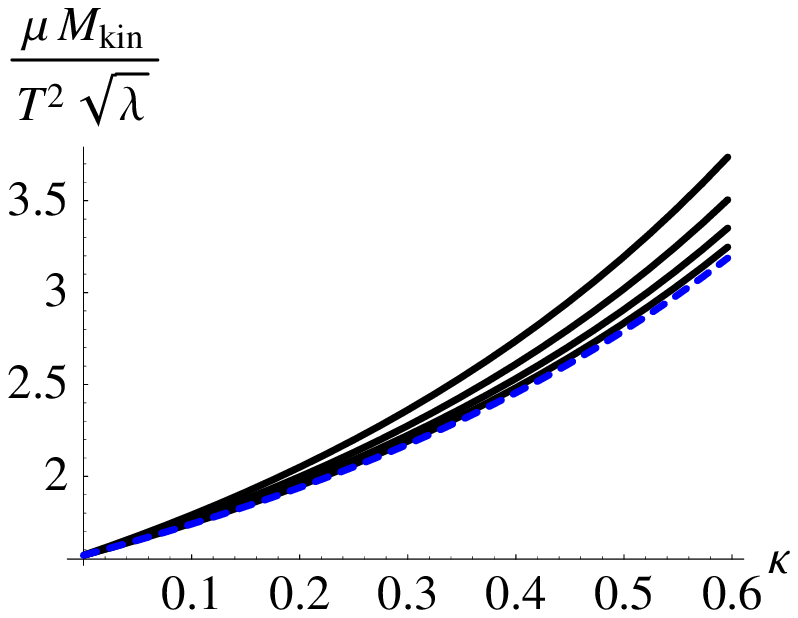, width=3in}}
\caption{The friction coefficient as a function of $\kappa$ for different values of $v$. 
The bottom most curve is the small $v$ limit.  As $v$ increases, the damping increases.
Also shown, from bottom to top, are the curves for $v=0.3$, 0.5, 0.7, and 0.9:
a) $\kappa_1=\kappa$ and $\kappa_2 = \kappa_3=0$,
b) $\kappa_1=\kappa_2 = \kappa$ and $\kappa_3=0$,
c) $\kappa_1=\kappa_2 = \kappa_3=\kappa$.
}
\label{vdep}
}

Although the formulae are messy, one can find explicit expressions for the integration constant 
$C$ of the analytic, constant velocity solution and hence for the energy and momentum loss.  For example, in the case $\kappa_1=\kappa_2=\kappa$ and $\kappa_3 = 0$, one finds
\be
 \frac{\mu \Mkin }{T^2 \sqrt{\lambda}} 
 =
 \frac{\pi}{2}
 \left( 
\frac{(1+\kappa)^2(1+\kappa(2 + v^2 \kappa))}{1+\kappa + \sqrt{1-v^2} \kappa}
\right)^{1/3} \ .
\label{mutwo}
\ee
This expression has the small $\kappa$ expansion
\be
 \frac{\mu \Mkin }{T^2 \sqrt{\lambda}} 
 =
 \frac{\pi}{2}
   \left( 1+ \frac{1}{3} (3-\sqrt{1-v^2}) \kappa + {\mathcal O}(\kappa^2) 
\right) \ .
\ee
and the small $v$ expansion
\be
 \frac{\mu \Mkin }{T^2 \sqrt{\lambda}} 
 =
 \frac{\pi}{2}  (1 + \kappa)^{2/3}  \left( 1+ \frac{1}{6} \kappa v^2 + {\mathcal O}(v^4) \right) \ .
\ee

Another simpler case is $\kappa_1=\kappa$ and $\kappa_2=\kappa_3=0$, for which we find
\be
 \frac{\mu \Mkin }{T^2 \sqrt{\lambda}} 
 =
\frac{2^{1/6} 2 \pi(1+\kappa)^{5/3} }{(2+\kappa)^{2}\left(2+2\kappa +\kappa^2- v^2 \kappa^2 + \kappa \sqrt{(1-v^2)( (2+\kappa)^2 - v^2 \kappa^2)}\right)^{1/6}}  \ .
\label{muone}
\ee
This expression has the small $\kappa$ expansion
\be
 \frac{\mu \Mkin }{T^2 \sqrt{\lambda}} 
 =
 \frac{\pi}{2}
\left( 1 + \frac{1}{6}(3-\sqrt{1-v^2}) \kappa + {\mathcal O}(\kappa^2) \right) \ ,
\ee
and the small $v$ expansion
\be
 \frac{\mu \Mkin }{T^2 \sqrt{\lambda}} 
 =
2 \pi \frac{(1 + \kappa)^{4/3} }{(2+\kappa)^2} \left(1 
+
\frac{\kappa}{6(2+\kappa)} v^2 + {\mathcal O}(v^4) \right) \ .
\ee

While the $\kappa_i=0$ result for $\mu \Mkin$ is independent of the velocity, interestingly,
nonzero chemical potential introduces a nontrivial dependence of $\mu$ on $v$.
As $v$ increases, as is clear from Figure \ref{vdep}, $\mu$ increases.

Before moving on to an analysis of the quasinormal modes for our string, we consider
the relativistic limit $v \to 1$ of (\ref{mutwo}) and (\ref{muone}).  Both of these expressions
for $\mu \Mkin$ approach a finite limit as $v \to 1$.  In the case $\kappa_1 = \kappa_2 =\kappa$ and $\kappa_3 = 0$, we find
\be
 \frac{\mu \Mkin }{T^2 \sqrt{\lambda}} 
 = \frac{\pi}{2} (1 + \kappa) \left( 1 - \frac{\kappa \sqrt{1-v^2}}{3(1+\kappa)}  + {\mathcal O}(1-v^2) \right)
\ee
while in the case $\kappa_1 = \kappa$ and $\kappa_2 = \kappa_3 = 0$, we get
\be
 \frac{\mu \Mkin }{T^2 \sqrt{\lambda}}  = \frac{2 \pi (1+ \kappa)^{3/2}}{(2+\kappa)^2}
 \left( 
 1 - \frac{\kappa \sqrt{1-v^2}}{6\sqrt{1+\kappa}}  + {\mathcal O}(1-v^2) \
 \right) \ .
\ee
Although $\mu \Mkin$ is finite in the relativistic limit, the derivative of $\mu$ with respect
to $v$ diverges at $v=1$ for nonzero chemical potential.

\subsection{Quasinormal Modes}
\label{sec:numerics}

We were not able to solve analytically the linearized equation of motion (\ref{lineom})
for the string in this black hole background.  However, we were able to
find $\mu$ as a function of $u_0$ numerically.  A simple shooting algorithm suffices.
At a point close to the horizon $u=(1 + \epsilon) u_h$, we use 
(\ref{impasymp}) to enforce the out-going boundary conditions.  For various
values of $\mu$ we integrate (\ref{lineom}) out to the flavor brane $u=u_0$.
By refining the choice of $\mu$, we locate the value that satisfies Neumann
boundary conditions $x'(u_0) = 0$. 

In Figure \ref{qnmplot}, we plot the friction coefficient as a function of $\Mrest$ for various
choices of $\kappa_i$.  We have introduced $\Delta m = \sqrt{\lambda} T/2$ to plot
a dimensionless quantity for $\Mrest$.
As predicted from the analysis of Section \ref{sec:linear},
$\mu \to 2 \pi T$ in the limit $u_0 \to u_h$.  The plots are also consistent with
the prediction that $\mu$ scales as $1/u_0$ in the large $u_0$ limit.
In between these two limits, $\mu$ is a monotone decreasing function of $u_0$,
lending credence to our hypothesis that $\mu$ is bounded above by $2 \pi T$ for small $v$.

\FIGURE{
\centerline{\raisebox{3cm}{(a)} \psfig{figure=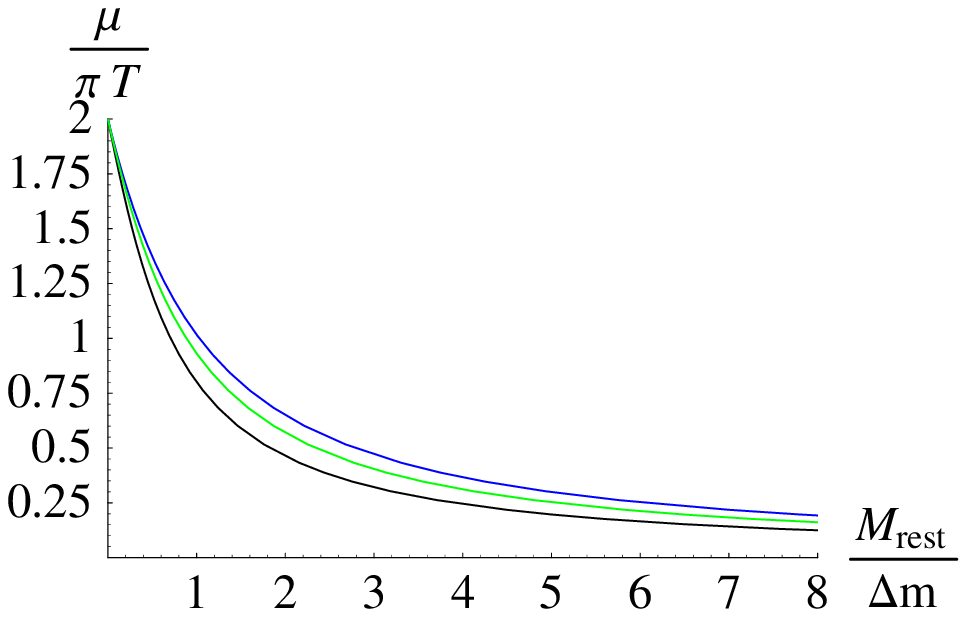, width=3in}
\raisebox{3cm}{(b)}
 \psfig{figure=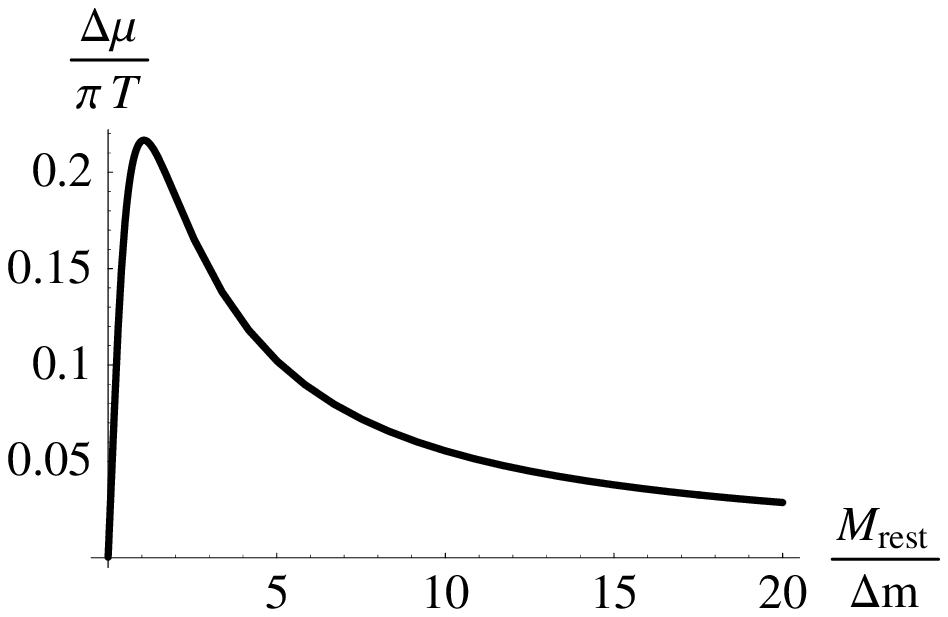, width=3in}}
\centerline{\raisebox{3cm}{(c)} \psfig{figure=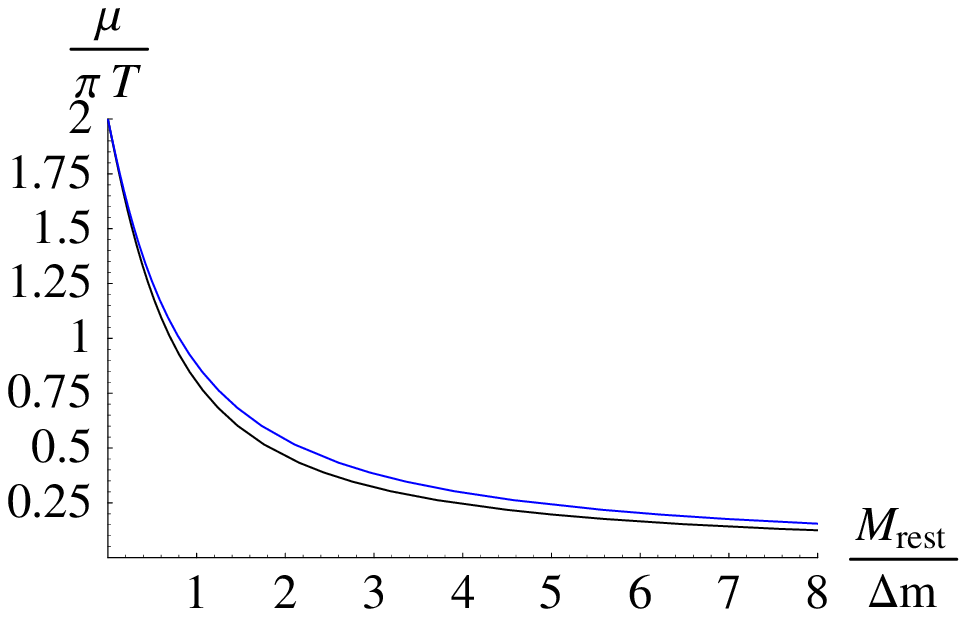, width=3in}
\raisebox{3cm}{(d)} \psfig{figure=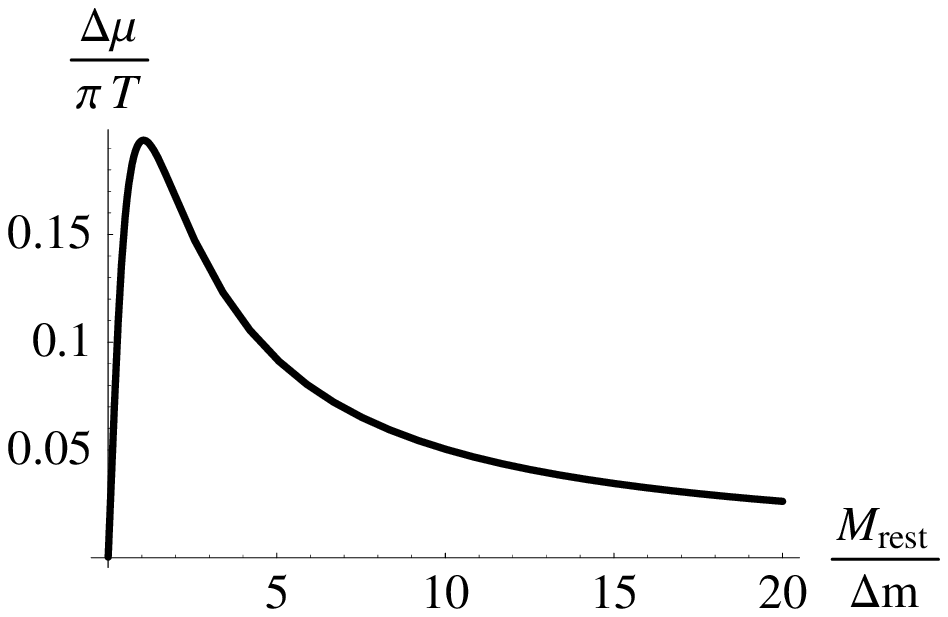, width=3in}}
\caption{The friction coefficient as a function of $\Mrest$
for different values of $\kappa$.
Plots (a) and (b) have $\kappa=\kappa_1=\kappa_2$ and $\kappa_3=0$.
Plots (c) and (d) have $\kappa=\kappa_1=\kappa_2=\kappa_3$.
Plot (a) shows from bottom to top $\kappa=0$, 1/2, and 1.  Plot (c)
shows from bottom to top $\kappa=0$ and $\kappa=0.596$.  
Plots (b) and (d) are the difference between the top and bottom
curve in (a) and (c) respectively.}
\label{qnmplot}
}

\FIGURE{
\centerline{\raisebox{3cm}{(a)} \psfig{figure=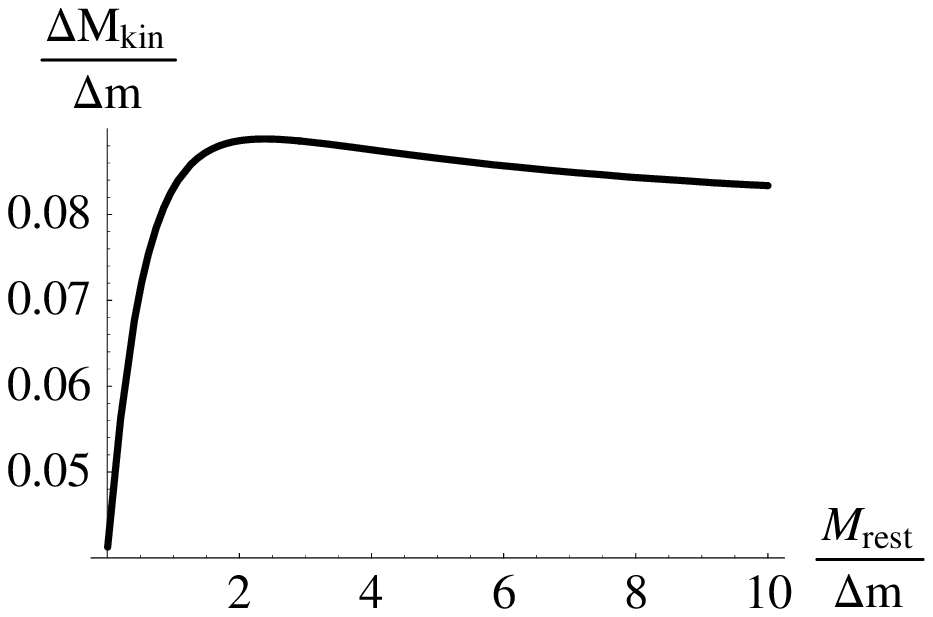, width=3in}
\raisebox{3cm}{(b)} \psfig{figure=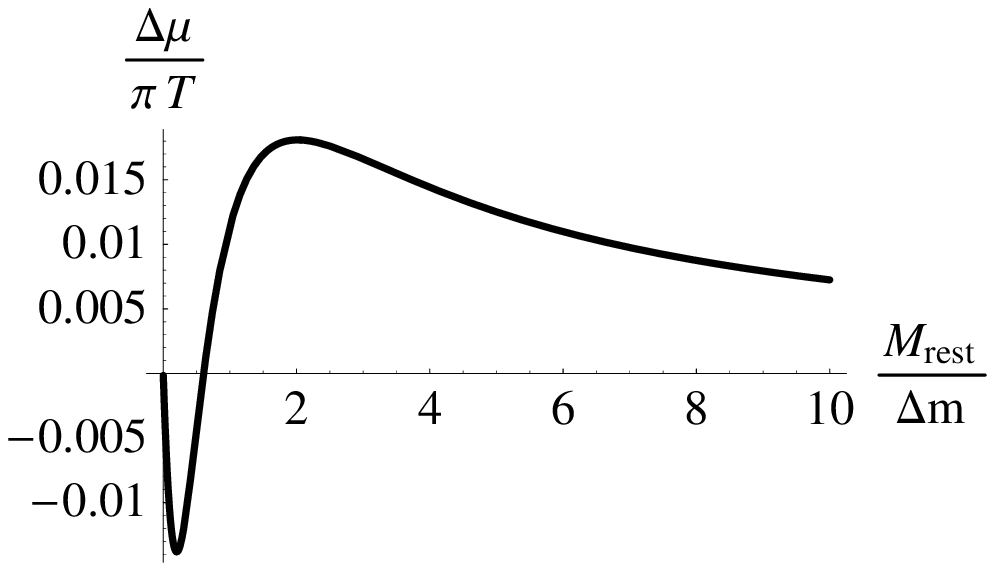, width=3in}}
\caption{Here $\kappa_2=\kappa_3=0$ and $\kappa_1=\kappa$.
Plot (a) is the difference $\Mkin(\kappa{=}2) - \Mkin(\kappa{=}0)$ 
as a function of $\Mrest$.
Plot (b) is the difference in $\mu$ for the same choices of $\kappa$.   
}
\label{remainingcase}
}

\section*{Acknowledgments}
I would like to thank  C.~Kozcaz and D.~Marolf for useful discussions.
Special thanks go to A.~Karch and L.~Yaffe
 for their encouragement and comments on the manuscript.
This work was supported in part by the U.S. Department
    of Energy under Grant No.~DE-FG02-96ER40956.

\bibliography{gkp}

\providecommand{\href}[2]{#2}\begingroup\raggedright\begin{thebibliography}{10}

\bibitem{HKKKY}
C.~P. Herzog, A.~Karch, P.~Kovtun, C.~Kozcaz, and L.~G. Yaffe, {\it Energy loss
  of a heavy quark in {N}=4 super {Y}ang-{M}ills plasma},
  \href{http://xxx.lanl.gov/abs/hep-th/0605158}{{\tt hep-th/0605158}}.

\bibitem{jthroat}
J.~M. Maldacena, {\it The large {$N$} limit of superconformal field theories
  and supergravity},  {\em Adv. Theor. Math. Phys.} {\bf 2} (1998) 231--252,
  [\href{http://xxx.lanl.gov/abs/hep-th/9711200}{{\tt hep-th/9711200}}].

\bibitem{GKP}
S.~S. Gubser, I.~R. Klebanov, and A.~M. Polyakov, {\it Gauge theory correlators
  from non-critical string theory},  {\em Phys. Lett.} {\bf B428} (1998)
  105--114, [\href{http://xxx.lanl.gov/abs/hep-th/9802109}{{\tt
  hep-th/9802109}}].

\bibitem{EW}
E.~Witten, {\it Anti-de {S}itter space and holography},  {\em Adv. Theor. Math.
  Phys.} {\bf 2} (1998) 253--291,
  [\href{http://xxx.lanl.gov/abs/hep-th/9802150}{{\tt hep-th/9802150}}].

\bibitem{Rajagopal}
H.~Liu, K.~Rajagopal, and U.~A. Wiedemann, {\it Calculating the jet quenching
  parameter from {AdS/CFT}},
  \href{http://xxx.lanl.gov/abs/hep-ph/0605178}{{\tt hep-ph/0605178}}.

\bibitem{SolanaTeaney}
J.~Casalderrey-Solana and D.~Teaney, {\it Heavy quark diffusion in strongly
  coupled {N}=4 {Y}ang {M}ills},
  \href{http://xxx.lanl.gov/abs/hep-ph/0605199}{{\tt hep-ph/0605199}}.

\bibitem{Gubser}
S.~S. Gubser, {\it Drag force in {AdS/CFT}},
  \href{http://xxx.lanl.gov/abs/hep-th/0605182}{{\tt hep-th/0605182}}.

\bibitem{Buchelnew}
A.~Buchel, {\it On jet quenching parameters in strongly coupled non-conformal
  gauge theories},  \href{http://xxx.lanl.gov/abs/hep-th/0605178}{{\tt
  hep-th/0605178}}.

\bibitem{Caceres}
E.~Caceres and A.~Guijosa, {\it Drag force in charged {N}=4 {SYM} plasma},
  \href{http://xxx.lanl.gov/abs/hep-th/0605235}{{\tt hep-th/0605235}}.

\bibitem{Adler:2005xv}
{\bf PHENIX} Collaboration, S.~S. Adler {\em et.~al.}, {\it Nuclear
  modification of electron spectra and implications for heavy quark energy loss
  in {A}u + {A}u collisions at {$s(\mbox {NN})^{1/2} = 200$ GeV}},  {\em
  Phys.~Rev.~Lett.} {\bf 96} (2006) 032301,
  [\href{http://xxx.lanl.gov/abs/nucl-ex/0510047}{{\tt nucl-ex/0510047}}].

\bibitem{CalderondelaBarcaSanchez:2005yp}
{\bf STAR} Collaboration, M.~Calderon de~la Barca~Sanchez {\em et.~al.}, {\it
  Open charm production from d + au collisions in star},  {\em Eur. Phys. J.}
  {\bf C43} (2005) 187--192.

\bibitem{Suaide:2005yq}
{\bf STAR} Collaboration, A.~A.~P. Suaide {\em et.~al.}, {\it Charm production
  in the star experiment at rhic},  {\em Eur. Phys. J.} {\bf C43} (2005)
  193--200.

\bibitem{Witten:1998zw}
E.~Witten, {\it Anti-de {S}itter space, thermal phase transition, and
  confinement in gauge theories},  {\em Adv. Theor. Math. Phys.} {\bf 2} (1998)
  505--532, [\href{http://xxx.lanl.gov/abs/hep-th/9803131}{{\tt
  hep-th/9803131}}].

\bibitem{Gubser:1998jb}
S.~S. Gubser, {\it Thermodynamics of spinning {D3}-branes},  {\em Nucl. Phys.}
  {\bf B551} (1999) 667--684,
  [\href{http://xxx.lanl.gov/abs/hep-th/9810225}{{\tt hep-th/9810225}}].

\bibitem{Behrndt}
K.~Behrndt, M.~Cvetic, and W.~A. Sabra, {\it Non-extreme black holes of five
  dimensional n = 2 ads supergravity},  {\em Nucl. Phys.} {\bf B553} (1999)
  317--332, [\href{http://xxx.lanl.gov/abs/hep-th/9810227}{{\tt
  hep-th/9810227}}].

\bibitem{Cai:1998ji}
R.-G. Cai and K.-S. Soh, {\it Critical behavior in the rotating {D}-branes},
  {\em Mod. Phys. Lett.} {\bf A14} (1999) 1895--1908,
  [\href{http://xxx.lanl.gov/abs/hep-th/9812121}{{\tt hep-th/9812121}}].

\bibitem{Chamblin:1999tk}
A.~Chamblin, R.~Emparan, C.~V. Johnson, and R.~C. Myers, {\it Charged {AdS}
  black holes and catastrophic holography},  {\em Phys. Rev.} {\bf D60} (1999)
  064018, [\href{http://xxx.lanl.gov/abs/hep-th/9902170}{{\tt
  hep-th/9902170}}].

\bibitem{CveticGubser}
M.~Cvetic and S.~S. Gubser, {\it Phases of r-charged black holes, spinning
  branes and strongly coupled gauge theories},  {\em JHEP} {\bf 04} (1999) 024,
  [\href{http://xxx.lanl.gov/abs/hep-th/9902195}{{\tt hep-th/9902195}}].

\bibitem{KarchKatz}
A.~Karch and E.~Katz, {\it Adding flavor to {AdS/CFT}},  {\em JHEP} {\bf 06}
  (2002) 043, [\href{http://xxx.lanl.gov/abs/hep-th/0205236}{{\tt
  hep-th/0205236}}].

\bibitem{GubserF}
S.~S. Gubser, I.~R. Klebanov, and A.~W. Peet, {\it Entropy and temperature of
  black 3-branes},  {\em Phys. Rev.} {\bf D54} (1996) 3915--3919,
  [\href{http://xxx.lanl.gov/abs/hep-th/9602135}{{\tt hep-th/9602135}}].

\bibitem{kovtun}
P.~Kovtun, D.~T. Son, and A.~O. Starinets, {\it Holography and hydrodynamics:
  Diffusion on stretched horizons},  {\em JHEP} {\bf 10} (2003) 064,
  [\href{http://xxx.lanl.gov/abs/hep-th/0309213}{{\tt hep-th/0309213}}].

\bibitem{Buchel}
A.~Buchel, {\it On universality of stress-energy tensor correlation functions
  in supergravity},  {\em Phys. Lett.} {\bf B609} (2005) 392--401,
  [\href{http://xxx.lanl.gov/abs/hep-th/0408095}{{\tt hep-th/0408095}}].

\bibitem{Kovtun:2004de}
P.~Kovtun, D.~T. Son, and A.~O. Starinets, {\it Viscosity in strongly
  interacting quantum field theories from black hole physics},  {\em Phys. Rev.
  Lett.} {\bf 94} (2005) 111601,
  [\href{http://xxx.lanl.gov/abs/hep-th/0405231}{{\tt hep-th/0405231}}].

\bibitem{Shuryak}
E.~Shuryak, {\it Why does the quark gluon plasma at {RHIC} behave as a nearly
  ideal fluid?},  {\em Prog. Part. Nucl. Phys.} {\bf 53} (2004) 273--303,
  [\href{http://xxx.lanl.gov/abs/hep-ph/0312227}{{\tt hep-ph/0312227}}].

\bibitem{Shuryak:2004cy}
E.~V. Shuryak, {\it What {RHIC} experiments and theory tell us about properties
  of quark-gluon plasma?},  {\em Nucl. Phys.} {\bf A750} (2005) 64--83,
  [\href{http://xxx.lanl.gov/abs/hep-ph/0405066}{{\tt hep-ph/0405066}}].

\bibitem{Babington:2003vm}
J.~Babington, J.~Erdmenger, N.~J. Evans, Z.~Guralnik, and I.~Kirsch, {\it
  Chiral symmetry breaking and pions in non-supersymmetric gauge/gravity
  duals},  {\em Phys. Rev.} {\bf D69} (2004) 066007,
  [\href{http://xxx.lanl.gov/abs/hep-th/0306018}{{\tt hep-th/0306018}}].

\bibitem{andy}
A.~O'Bannon and A.~Karch, {\it Chiral transition of {N=4} super {Y}ang-{M}ills
  with flavor on a 3-sphere},
  \href{http://xxx.lanl.gov/abs/hep-th/0605120}{{\tt hep-th/0605120}}.

\bibitem{Mateos:2006nu}
D.~Mateos, R.~C. Myers, and R.~M. Thomson, {\it Holographic phase transitions
  with fundamental matter},  \href{http://xxx.lanl.gov/abs/hep-th/0605046}{{\tt
  hep-th/0605046}}.

\bibitem{Albash:2006ew}
T.~Albash, V.~Filev, C.~V. Johnson, and A.~Kundu, {\it A topology-changing
  phase transition and the dynamics of flavour},
  \href{http://xxx.lanl.gov/abs/hep-th/0605088}{{\tt hep-th/0605088}}.

\bibitem{AlbashRcharge}
T.~Albash, V.~Filev, C.~V. Johnson, and A.~Kundu, {\it Global currents, phase
  transitions and chiral symmetry breaking in large {N} gauge theory},
  \href{http://xxx.lanl.gov/abs/hep-th/0605175}{{\tt hep-th/0605175}}.

\bibitem{MooreTeaney}
G.~D. Moore and D.~Teaney, {\it How much do heavy quarks thermalize in a heavy
  ion collision?},  {\em Phys. Rev.} {\bf C71} (2005) 064904,
  [\href{http://xxx.lanl.gov/abs/hep-ph/0412346}{{\tt hep-ph/0412346}}].

\bibitem{SonStarinets}
D.~T. Son and A.~O. Starinets, {\it Minkowski-space correlators in {AdS/CFT}
  correspondence: Recipe and applications},  {\em JHEP} {\bf 09} (2002) 042,
  [\href{http://xxx.lanl.gov/abs/hep-th/0205051}{{\tt hep-th/0205051}}].

\end{thebibliography}\endgroup
\bibliographystyle{JHEP}

\end{document}